\documentclass[conference]{IEEEtran}

\usepackage{graphicx}
\usepackage{multirow}
\usepackage{multicol}
\usepackage{threeparttablex}
\usepackage{pifont}
\usepackage{mathtools}
\usepackage{subfig}

\newcommand{\cmarkn}{\ding{51}}
\newcommand{\xmark}{\ding{55}}

\IEEEoverridecommandlockouts

\title{DNS based In-Browser Cryptojacking Detection}

\author{
    \IEEEauthorblockN{
        Rohit Kumar Sachan\IEEEauthorrefmark{1},
        Rachit Agarwal\IEEEauthorrefmark{2},
        Sandeep Kumar Shukla\IEEEauthorrefmark{2}
    }
    \IEEEauthorblockA{
        \IEEEauthorrefmark{1}C3i Hub, IIT Kanpur, India
        \IEEEauthorrefmark{2}CSE, IIT Kanpur, India\\
        Email: \IEEEauthorrefmark{1}{sachan.rohit100@gmail.com},
        \IEEEauthorrefmark{2}{\{rachitag, sandeeps\}}@cse.iitk.ac.in
    }
}

\begin{document}

\maketitle

\begin{abstract}
The metadata aspect of Domain Names (DNs) enables us to perform a behavioral study of DNs and detect if a DN is involved in in-browser cryptojacking. Thus, we are motivated to study different temporal and behavioral aspects of DNs involved in cryptojacking. We use temporal features such as query frequency and query burst along with graph-based features such as degree and diameter, and non-temporal features such as the string-based to detect if a DNs is suspect to be involved in the in-browser cryptojacking. Then, we use them to train the Machine Learning (ML) algorithms over different temporal granularities such as 2 hours datasets and complete dataset. Our results show DecisionTrees classifier performs the best with 59.5\% Recall on cryptojacked DN, while for unsupervised learning, K-Means with K=2 perform the best. Similarity analysis of the features reveals a minimal divergence between the cryptojacking DNs and other already known malicious DNs. It also reveals the need for improvements in the feature set of state-of-the-art methods to improve their accuracy in detecting in-browser cryptojacking. As added analysis, our signature-based analysis identifies that none-of-the Indian Government websites were involved in cryptojacking during October-December 2021. However, based on the resource utilization, we identify 10 DNs with different properties than others.

\end{abstract}


\section{Introduction}

\textbf{\textit{Cryptojacking}} is a distributed mining approach in which cyber-criminals perform cryptocurrency mining activities illegally over the Internet by infecting a user's device. Here, Crypto-miners illegally control the user's device computational resources for cryptocurrency mining purposes either by \textit{(a)} installing malware that performs mining activities or \textit{(b)} when a user visits some URL/website, till the time user is active on the URL, in the background execute mining scripts on the user devices. Such techniques facilitate the miners to get financial benefits without compromising their computational resources, cost, and sharing of the mining rewards with the compromised user.
Ameliorated with sanctions of a state on the mining processes, environmental concerns~\cite{miningbanned}, and the adoption of cryptocurrencies, cryptojacking is increasing at an alarming pace and becoming a concern for cyber security experts~\cite{spikecrypto, risecrypto}. To limit cryptojacking, some security companies such as Norton have started to provide mining pool services to their users officially. 

Besides cryptojacking malware, one of the most common and easy ways to perform cryptojacking is via \textbf{\textit{in-browser cryptojacking}}. Here, cyber-criminals use \textit{JavaScript (JS)} and \textit{WebAssembly (WASM)} to perform cryptojacking. Such a type of cryptojacking is easy to implement but hard to detect. Now defunct, one of the most popular mining scripts was developed by CoinHive~\cite{coinhive}. Many website owners (especially those involved in gaming and multimedia content) use such mining scripts on their websites for alternate revenue~\cite{krebs2018and}. In~\cite{varlioglu2020cryptojacking}, the authors investigated if the discontinuation of CoinHive impacted cryptojacking. Using CMTracker~\cite{hong2018you}, they concluded that attackers evolved and introduced new mining scripts. Thus, it does not impact cryptojacking.

In-browser cryptojacking detection techniques mainly crawl the source code of the websites to extract explicit keywords or signatures~\cite{konoth2018minesweeper, bijmans2019inadvertently}. Further, some techniques analyse \textit{(a)} computational resource utilization (CPU, GPU, memory, disk)~\cite{konoth2018minesweeper, ning2019capjack}, \textit{(b)} scripting code~\cite{konoth2018minesweeper, bursztein2020coinpolice}, \textit{(c)} opcode~\cite{carlin2018detecting, naseem2021minos}, \textit{(d)} trace network packets~\cite{saad2018end, saad2019dine}, and \textit{(e)} hash function~\cite{hong2018you, romano2020minerray} of mining script. On the contrary, for evasion, cryptojackers now use different techniques such as \textit{CPU limiting, code obfuscation, payload hiding, and changing the used script frequently} to evade naive detection approaches. 

Websites also have a unique signature on their metadata. Such metadata includes Domain Name (DN) and Domain Name System (DNS) records (including IP address, NS address, location, and others). Thus, \textbf{\textit{can such metadata (DN and DNS records) help detect websites performing/involved in in-browser cryptojacking?}} In one of the state-of-the-art approaches~\cite{sachan2021identifying} (for details, refer to Section~\ref{sec:METH}), the authors presented an approach to detect suspicious domains using temporal and non-temporal properties of DNS records in the blockchain ecosystem. They analyzed the DNS traffic records and identified temporal (i.e., time-series based) and non-temporal (i.e., non-time series based) properties to understand the actual behavior of DNs on two temporal granularities (i.e., 2H (sub-datasets of 2 hour duration) and ALL (complete dataset)). As in-browser cryptojacking is one type of malicious/illicit activity, it motivates us to check if the approaches such as~\cite{sachan2021identifying} can be used to detect in-browser cryptojacking. Here, we check the impact of the metadata information on the detection of cryptojacking websites in two ways, \textit{(i)} we study and analyze the similarity between the features of in-browser cryptojacking DNs and other malicious DNs and \textit{(ii)} we validate if existing state-of-the-art methods can detect the in-browser cryptojacking. 
We observe DecisionTrees classifier performs the best with 59.5\% Recall among other supervised ML algorithms, and 228 DNs show high similarity with malicious DNs across different temporal granularities using K-Mean with K=2.

Further, in the past Indian Government websites have witnessed in-browser cryptojacking~\cite{reportET2018}. Thus, apart from the above validations, we also perform \textit{\textbf{an analysis of Indian Government websites from the cryptojacking perspective}} to know  if any Indian Government website is under attack. This analysis includes signature crawling and resource utilization analysis (i.e., CPU, device, disk, and network) and whois record. Here, we perform K-Mean clustering (because of the unavailability of ground truth) using resource utilization features to identify the DNs with distinct resource utilization. Note that we understand that cryptojacking is dynamic (source code of websites may change over time), and Wayback Machine archives may provide old snap-shots of source codes. Still, Wayback Machine does not log associated scripting codes, which is essential to us. Due to this unavailability of associated scripting codes and associated DNS for all Indian Government websites, we cannot use any state-of-the-art method such as~\cite{sachan2021identifying} for the analysis. Our analysis reveals that none-of-the website contains a cryptojacking signature in their code. Most of the websites are clustered in one cluster based on our feature set (based on resource utilization, cf. Table~\ref{table:4}). 
Our analysis also identifies the distinct resource utilization by 10 Indian DNs, which should be investigated further. 
From this point forward, we refer to in-browser cryptojacking as cryptojacking interchangeably.

In short, our main contributions are:

\begin{itemize}
    \item \textbf{Comparative study:} We present a comparative study of the various state-of-the-art techniques used to detect in-browser cryptojacking. Here, we compare these state-of-the-art techniques based on the features used, classifier/method, dataset with the size, reported performance, and limitations. We identify that no technique uses DNS records for the in-browser cryptojacking detection.
    
    \item \textbf{Similarity analysis between cryptojacking DNs and other malicious DNs} 
    revealed the minimal divergence between temporal features of malicious DNs and cryptojacking DNs. 
    
    \item \textbf{Effectiveness of the state-of-the-art methods such as~\cite{sachan2021identifying} towards identifying cryptojacked DNs:} Our validations reveal the need for improvement in the feature set of the state-of-the-art methods such as~\cite{sachan2021identifying} to improve their performance in detecting in-browser cryptojacking. 
    
    \item \textbf{Analysis of Indian Government websites} reveals that none-of-the Indian Government websites were involved in cryptojacking during October-December 2021, and the distinct resource utilization by 10 Indian Government DNs.

\end{itemize}

This paper is structured as follows. The state-of-the-art associated with the detection of cryptojacking is presented in Section~\ref{sec:RW}. Section~\ref{sec:METH} presents our methodology with an in-depth validation accompanied by the result analysis in Section~\ref{sec:results}. Finally, we conclude with Section~\ref{sec:conc}.

\begin{table*}
\centering
\caption{Summary of related studies}
\label{table:rw}
\begin{threeparttable}
\scriptsize{
\begin{tabular}{|c|c|p{0.05cm}p{0.05cm}p{0.05cm}p{0.05cm}p{0.05cm}p{0.05cm}p{0.05cm}p{0.05cm}p{0.2cm}p{0.2cm}|c|c|c|c|c|}
    \hline
     
    \multicolumn{2}{|c|}{Technique} &
    \multicolumn{10}{c|}{Based On} & 
    \multirow{2}{*}{Method} &  
    \multicolumn{2}{c|}{Datasets} &
    Performance / & 
    \multirow{2}{*}{Limitation}\\
    \cline{1-2} \cline{3-12} \cline{14-15}
    & Ref. & S & P & M & D & N & C & O & H & DNS & Oth &  & Source & Size & Results & \\
    \hline

    \multirow{3}{*}{\rotatebox[origin=c]{90}{Static}} &
    \multirow{3}{*}{\cite{romano2020minerray}} &
    \multirow{3}{*}{\xmark} &
    \multirow{3}{*}{\xmark} &
    \multirow{3}{*}{\xmark} & 
    \multirow{3}{*}{\xmark} & 
    \multirow{3}{*}{\xmark} & 
    \multirow{3}{*}{\xmark} & 
    \multirow{3}{*}{\cmarkn} & 
    \multirow{3}{*}{\cmarkn} & 
    \multirow{3}{*}{\xmark} &
    \multirow{3}{*}{\xmark} & 
    \multirow{3}{*}{Crawling} & 
    \multirow{3}{*}{Alexa} & 
    \multirow{3}{*}{1.2M} & 
    \multirow{3}{*}{901 TLDs} & 
    \multirow{1}{*}{Unable to handle } \\
    & & & & & & & & & & & & & & & & obfuscation techniques \\
    & & & & & & & & & & & & & & & & and Memory overhead \\
    \hline
   
    \multirow{29}{*}{\rotatebox[origin=c]{90}{Dynamic}} &
    \multirow{2}{*}{\cite{hong2018you}} & 
    \multirow{2}{*}{\xmark} & 
    \multirow{2}{*}{\xmark} & 
    \multirow{2}{*}{\xmark} & 
    \multirow{2}{*}{\xmark} & 
    \multirow{2}{*}{\xmark} & 
    \multirow{2}{*}{\xmark} & 
    \multirow{2}{*}{\xmark} &
    \multirow{2}{*}{\cmarkn} & 
    \multirow{2}{*}{\xmark} & 
    \multirow{2}{*}{\cmarkn} & 
    Threshold- & 
    \multirow{2}{*}{Alexa} &
    \multirow{2}{*}{853K} &
    \multirow{2}{*}{2770 TLDs} & 
    \multirow{1}{*}{Detects only hash} \\
    & & & & & & & & & & & & based& & & & modeled signatures \\
    \cline{2-17}

    &
    \multirow{5}{*}{\cite{carlin2018detecting}} & 
    \multirow{5}{*}{\xmark} &
    \multirow{5}{*}{\xmark} &
    \multirow{5}{*}{\xmark} &
    \multirow{5}{*}{\xmark} & 
    \multirow{5}{*}{\xmark} & 
    \multirow{5}{*}{\xmark} & 
    \multirow{5}{*}{\cmarkn} & 
    \multirow{5}{*}{\xmark} & 
    \multirow{5}{*}{\xmark} &
    \multirow{5}{*}{\xmark} & 
    \multirow{5}{*}{RF} & 
    \multirow{1}{*}{} & 
    \multirow{5}{*}{1K} & 
    \multirow{1}{*}{Acc=$>$99.0\% } & 
    \multirow{1}{*}{} \\
    & & & & & & & & & & & & & VirusShare & & Recall=99.2\% & Performance validated \\
    & & & & & & & & & & & & & OpenDNS & & Precision=99.2\% & on limited data  \\
    & & & & & & & & & & & & & & & TPR=99.2\% & \\
    & & & & & & & & & & & & & & & FPR=0.9\% & \\
    \cline{2-17}

    &
    \multirow{7}{*}{\cite{papadopoulos2019truth}} &
    \multirow{7}{*}{\xmark} &
    \multirow{7}{*}{\cmarkn} &
    \multirow{7}{*}{\cmarkn} & 
    \multirow{7}{*}{\xmark} & 
    \multirow{7}{*}{\cmarkn} & 
    \multirow{7}{*}{\xmark} & 
    \multirow{7}{*}{\xmark} & 
    \multirow{7}{*}{\xmark} & 
    \multirow{7}{*}{\xmark} &
    \multirow{7}{*}{\cmarkn} & 
    \multirow{7}{*}{Crawling} & 
    \multirow{1}{*}{Alexa} & 
    \multirow{7}{*}{200K} & 
    \multirow{1}{*}{} & 
    \multirow{1}{*}{} \\
    & & & & & & & & & & & & & BlackLists, & & Profit$\approx$5.5$\times\downarrow$ & \\
    & & & & & & & & & & & & & PublicWWW, & & CPU$\approx$59$\times\uparrow$ & Performance \\
    & & & & & & & & & & & & & CoinHive,  & & Temp$\approx$52.8$\times\uparrow$ & and Time \\
    & & & & & & & & & & & & & CryptoLoot,  & & Power$\approx$2.0$\times\uparrow$ & overhead \\
    & & & & & & & & & & & & & JSEcoin, & & & \\
    & & & & & & & & & & & & & CoinHave & & & \\
    \cline{2-17}

    &
    \multirow{3}{*}{\cite{bursztein2020coinpolice}} &
    \multirow{3}{*}{\xmark} &
    \multirow{3}{*}{\cmarkn} &
    \multirow{3}{*}{\xmark} &
    \multirow{3}{*}{\xmark} & 
    \multirow{3}{*}{\xmark} & 
    \multirow{3}{*}{\cmarkn} & 
    \multirow{3}{*}{\xmark} & 
    \multirow{3}{*}{\xmark} & 
    \multirow{3}{*}{\xmark} &
    \multirow{3}{*}{\cmarkn} & 
    \multirow{3}{*}{CNN} & 
    \multirow{3}{*}{Alexa} & 
    \multirow{3}{*}{47K} & 
    \multirow{1}{*}{Acc=98.7\%} & 
    \multirow{1}{*}{Address exclusively} \\
    & & & & & & & & & & & & & & & TPR=97.87\% & browser-based mining \\
    & & & & & & & & & & & & & & & FPR=0.74\% & \\
    \cline{2-17}

    &
    \multirow{4}{*}{~\cite{gomes2020cryptojacking}} & 
    \multirow{4}{*}{\xmark} &
    \multirow{4}{*}{\cmarkn} &
    \multirow{4}{*}{\xmark} & 
    \multirow{4}{*}{\xmark} & 
    \multirow{4}{*}{\xmark} & 
    \multirow{4}{*}{\xmark} & 
    \multirow{4}{*}{\xmark} & 
    \multirow{4}{*}{\xmark} &
    \multirow{4}{*}{\xmark} &
    \multirow{4}{*}{\xmark} & 
    \multirow{1}{*}{TLC, SMO,} & 
    \multirow{4}{*}{Alexa} & 
    \multirow{4}{*}{1.2K} & 
    \multirow{1}{*}{1837 TLDs} & 
    \multirow{1}{*}{Performance validated } \\
    & & & & & & & & & & & & MISVM, & & & Precision=1.0\% & on limited data \\
    & & & & & & & & & & & & Random & & & Recall=1.0\% & \\
    & & & & & & & & & & & & SubSpace & & & & \\
    \cline{2-17}

    &
    \multirow{3}{*}{\cite{gomes2020cryingjackpot}} & 
    \multirow{3}{*}{\xmark} &
    \multirow{3}{*}{\cmarkn} &
    \multirow{3}{*}{\cmarkn} & 
    \multirow{3}{*}{\xmark} & 
    \multirow{3}{*}{\cmarkn} & 
    \multirow{3}{*}{\xmark} & 
    \multirow{3}{*}{\xmark} & 
    \multirow{3}{*}{\xmark} & 
    \multirow{3}{*}{\xmark} &
    \multirow{3}{*}{\xmark} & 
    \multirow{1}{*}{K-Means } & 
    \multirow{1}{*}{Hybrid dataset,} & 
    \multirow{3}{*}{-} & 
    \multirow{1}{*}{Precision, Recall,} & 
    \multirow{1}{*}{Limited} \\
    & & & & & & & & & & & & DBSCAN & CIC-IDS2018 & & F1-Score= & mining samples\\
    & & & & & & & & & & & & Agglomerative & & & $>$92.0 & \\
    \cline{2-17}

    &
    \multirow{2}{*}{\cite{caprolu2021cryptomining}} & 
    \multirow{2}{*}{\xmark} &
    \multirow{2}{*}{\xmark} &
    \multirow{2}{*}{\xmark} & 
    \multirow{2}{*}{\xmark} & 
    \multirow{2}{*}{\cmarkn} & 
    \multirow{2}{*}{\xmark} & 
    \multirow{2}{*}{\xmark} & 
    \multirow{2}{*}{\xmark} & 
    \multirow{2}{*}{\xmark} & 
    \multirow{2}{*}{\cmarkn} & 
    \multirow{2}{*}{RF} & 
    \multirow{1}{*}{Self} & 
    \multirow{2}{*}{-} & 
    \multirow{1}{*}{F1-Score=96.0\%} & 
    \multirow{1}{*}{Solely relying on} \\
    & & & & & & & & & & & & & Generated & & AUC=99.0\% & the network traffic \\
    \cline{2-17}

    & 
    \multirow{3}{*}{\cite{naseem2021minos}} &
    \multirow{3}{*}{\xmark} &
    \multirow{3}{*}{\xmark} &
    \multirow{3}{*}{\xmark} & 
    \multirow{3}{*}{\xmark} & 
    \multirow{3}{*}{\xmark} & 
    \multirow{3}{*}{\xmark} & 
    \multirow{3}{*}{\cmarkn} & 
    \multirow{3}{*}{\xmark} & 
    \multirow{3}{*}{\xmark} &
    \multirow{3}{*}{\xmark} & 
    \multirow{3}{*}{CNN} & 
    \multirow{3}{*}{PublicWWW} & 
    \multirow{3}{*}{-} & 
    \multirow{1}{*}{Acc=98.97\%} & 
    \multirow{1}{*}{Considers only WASM} \\
    & & & & & & & & & & & & & & & Precision=93.07\% & modules and does not  \\
    & & & & & & & & & & & & & & & F1-Score=95.04\% & support JS modules\\
    \hline

    \multirow{14}{*}{\rotatebox[origin=c]{90}{Hybrid}} &
    \multirow{3}{*}{\cite{konoth2018minesweeper}} & 
    \multirow{3}{*}{\cmarkn} &
    \multirow{3}{*}{\cmarkn} &
    \multirow{3}{*}{\cmarkn} & 
    \multirow{3}{*}{\xmark} & 
    \multirow{3}{*}{\cmarkn} & 
    \multirow{3}{*}{\cmarkn} & 
    \multirow{3}{*}{\xmark} & 
    \multirow{3}{*}{\xmark} & 
    \multirow{3}{*}{\xmark} &
    \multirow{3}{*}{\xmark} & 
    \multirow{3}{*}{Crawling} & 
    \multirow{3}{*}{Alexa} & 
    \multirow{3}{*}{1M} & 
    \multirow{3}{*}{-} & 
    \multirow{1}{*}{Detect only CryptoNight} \\
    & & & & & & & & & & & & & & & & miners, Do not support \\
    & & & & & & & & & & & & & & & & JS miners\\
    \cline{2-17}

    &
\cite{saad2018end} & 
    \multirow{3}{*}{\cmarkn} &
    \multirow{3}{*}{\cmarkn} &
    \multirow{3}{*}{\xmark} & 
    \multirow{3}{*}{\xmark} & 
    \multirow{3}{*}{\cmarkn} & 
    \multirow{3}{*}{\cmarkn} & 
    \multirow{3}{*}{\xmark} & 
    \multirow{3}{*}{\xmark} & 
    \multirow{3}{*}{\xmark} &
    \multirow{3}{*}{\cmarkn} & 
    \multirow{1}{*}{FCM} & 
    \multirow{1}{*}{Pixalate} & 
    \multirow{3}{*}{5.7K} & 
    \multirow{1}{*}{Acc=96.4\%} & 
    \multirow{1}{*}{Scalability issue, } \\
    &\cite{saad2019dine} & & & & & & & & & & & SVM & Netlab360 & & FPR=3.3\% & Code obfuscation and \\
    & & & & & & & & & & & & RF & & & FNR=3.7\% & WASM are not considered \\
    \cline{2-17}

    &
    \multirow{3}{*}{\cite{ning2019capjack}} & 
    \multirow{3}{*}{\cmarkn} &
    \multirow{3}{*}{\cmarkn} &
    \multirow{3}{*}{\cmarkn} & 
    \multirow{3}{*}{\cmarkn} & 
    \multirow{3}{*}{\cmarkn} & 
    \multirow{3}{*}{\xmark} & 
    \multirow{3}{*}{\xmark} & 
    \multirow{3}{*}{\xmark} & 
    \multirow{3}{*}{\xmark} &
    \multirow{3}{*}{\xmark} & 
    \multirow{3}{*}{CNN} & 
    \multirow{1}{*}{Self} & 
    \multirow{3}{*}{1.8K} & 
    \multirow{1}{*}{DR=87.0\%} & 
    \multirow{1}{*}{Address exclusively} \\
    & & & & & & & & & & & & & Generated & & DR=99.0\% & browser-based mining \\
    & & & & & & & & & & & & & & & (after 11 sec.) & \\
    \cline{2-17}

    &
    \multirow{4}{*}{\cite{bijmans2019inadvertently}} &
    \multirow{4}{*}{\cmarkn} &
    \multirow{4}{*}{\xmark} &
    \multirow{4}{*}{\xmark} &
    \multirow{4}{*}{\xmark} & 
    \multirow{4}{*}{\cmarkn} & 
    \multirow{4}{*}{\xmark} & 
    \multirow{4}{*}{\xmark} & 
    \multirow{4}{*}{\xmark} & 
    \multirow{4}{*}{\xmark} &
    \multirow{4}{*}{\xmark} & 
    \multirow{4}{*}{Crawling} & 
    \multirow{1}{*}{Alexa,} & 
    \multirow{2}{*}{1.8M} & 
    \multirow{2}{*}{204 Campaigns} & 
    \multirow{1}{*}{} Exclusively depends \\ 
    & & & & & & & & & & & & & Majestic, & & &  on vulnerabilities of \\ \cline{16-17}
    & & & & & & & & & & & & & PublicWWW, & \multirow{2}{*}{48.9M} & \multirow{2}{*}{1136 TLDs} & CMS providers- \\
    & & & & & & & & & & & & &\cite{Umbrella1m} & & & such as WordPress \\
    \hline
\end{tabular}
}
\begin{tablenotes}
    \item $\bullet$ \textbf{Based on}: $^{S}$ Signature, $^{P}$ Processor / CPU, $^{M}$ Memory, $^{D}$ Disk, $^{N}$ Network Analysis, $^{C}$ Code Analysis, $^{O}$ Opcode, $^{H}$ Hashing Algorithm, $^{DNS}$ Domain Name System, $^{Oth}$ Others, $\bullet$ \textbf{Method}: $^{RF}$ Random Forest, $^{CNN}$ Convolutional Neural Network, $^{TLC}$ Two-Level Classification, $^{FCM}$ Fuzzy C-Means, $^{MISVM}$ Multiple-Instance Support Vector Machine, $^{SMO}$ Sequential Minimal Optimization, $^{RandomSubSpace}$ Random Subspace Method, $\bullet$ $^\text{\xmark}$ not used, $^\text{\cmarkn}$ used, $^-$ no specific mention,$^{\times}$ times
\end{tablenotes}
\end{threeparttable}
\vspace{-0.44cm}
\end{table*}

\section{Related Work}\label{sec:RW}

This section presents the state-of-the-art works related to the detection of in-browser cryptojacking.
In~\cite{tekiner2021sok}, the authors presented a survey of the cryptojacking malware detection techniques and an overview of two cryptojacking datasets and 45 significant cryptojacking attack instances. They classified the related techniques as static, dynamic, and hybrid approaches. A static technique uses signature search (or crawling) of known malware’s signature in scripting code. It analyzes the script code, opcode (machine level binary code), and hash algorithm to detect cryptojacking. Static tools such as MinerRay~\cite{romano2020minerray} infer signatures of the hash function and use an intermediate representation (IR) of both JS, and WASM and inspect interactions between the client and cryptojacking module for detection. A static technique suffers from the obfuscated or unseen signatures problem and requires up-to-date signatures for detection.

Dynamic techniques analyze the computational resources (i.e., processor/CPU, memory, disk, power, and others) and network traffic. These techniques are robust against evasion techniques such as scripting code and throttling and can capture any behavioral changes~\cite{hong2018you}. In~\cite{gomes2020cryptojacking}, the authors proposed a CPU usage metrics-based detector. In contrast, in~\cite{gomes2020cryingjackpot}, the authors proposed an approach-based on the host performance counter-based features (i.e., CPU, memory, network usage, and running processes within a host) and network flow-based features (i.e., inbound/outbound flows from port 80 and 443 as Stratum mining protocol utilizes them). Another dynamic approach-based tool called WebTestbench~\cite{papadopoulos2019truth} uses system resources, energy consumption, network traffic, device temperature, and user experience. While other approaches such as~\cite{carlin2018detecting} analyze the CPU instruction during the opcode execution, and~\cite{bursztein2020coinpolice} analyze the execution patterns of JS and WASM code and CPU utilization for detection. Similarly, Crypto-Aegis~\cite{caprolu2021cryptomining} analyzes the network traces generated by the node of Bitcoin, Monero, and ByteCoin under (i) no VPN, (ii) Nord VPN, and (iii) Express VPN, to identify the cryptomining activities (such as pool mining, solo mining, and full active node). Apart from the aforementioned dynamic techniques, MINOS~\cite{naseem2021minos} uses image-based classification and deep learning techniques to distinguish between benign and cryptojacked (i.e., those that have WASM script) opcode. 

Hybrid approaches are more prominent than the static and the dynamic approaches. Among hybrid approaches, MineSweeper~\cite{konoth2018minesweeper} uses signature crawling, WebSocket traffic analysis, CPU usage analysis, code analysis of WebAssembly script, and memory cache events during the execution. While CapJack~\cite{ning2019capjack} uses the CapsNet (Capsule Network) technology to measure the abnormal resource utilization. Similarly, approaches in~\cite{saad2018end} and~\cite{saad2019dine} perform static analysis based on content-based, currency-based, and code-based, while dynamic analysis is based on CPU and battery consumption. A content-based analysis is used to find the nature of websites such as entertainment and sports; a currency-based analysis is used to find the type of cryptocurrencies being mined through in-browser cryptojacking. In contrast, the code-based analysis is based on the code complexity of the script. Here, the code analysis includes cyclomatic complexity, cyclomatic complexity density, Halstead complexity, line of code, and maintainability score. The approach's results reveal 10-20 times higher CPU usage and $\approx$8 times more battery drainage by cryptojacking scripts. Further, in~\cite{bijmans2019inadvertently}, the authors introduced a cryptojacking campaigns detector based on the crawling and NetFlow data traffic. They used WebAssembly, asm.js (a technique translating high-level code, like C and C++ to JavaScript), WebSockets, and Stratum Mining Protocol to detect cryptojacking.

These state-of-the-art approaches are summarized in Table~\ref{table:rw} with the reported features, classifier/method, dataset used with the size, reported performance, and approach limitations. Our study identifies that none-of-the discussed approaches use DNS records/DNS-based techniques to detect in-browser cryptojacking.

\section{Methodology}\label{sec:METH}
Our approach follows the standard ML pipeline steps, including data collection, data pre-processing, feature engineering, ML algorithm, validation, and is motivated by~\cite{sachan2021identifying}, which identifies illicit DNs using temporal (i.e., time-series based) and non-temporal features.
The non-temporal features include the string-based and DNS query/response-based features. While the temporal features include the DNS query burst and DNS dynamic graph-based (such as degree and diameter) features. A burst is defined as an over-the-threshold value for a given feature. Consider a graph that links all IPs and NS addresses to the DN. A degree is thus defined as the number of edges (IPs and NS addresses) associated with a DN. Similarly, diameter is the largest shortest path of the graph component in which that DN exists. These temporal features are extracted using two temporal granularities: 2H (2 hours based data segments) and ALL (complete data). On these datasets, we then apply both supervised and unsupervised ML models to detect illicit DNs.

In the pre-processing step, we collect the data, label it (as benign, malicious, and cryptojacking) using publicly available sources, and segment it into different temporal granularities. Here, we extract all 48 temporal and non-temporal features (same as those in~\cite{sachan2021identifying}, due to space constraints, we do not list those features) and analyze the similarity (by comparing the probability distribution) between the temporal properties of the cryptojacking DNs and other malicious DNs. For the unsupervised ML, we first apply the reported unsupervised algorithm (as in~\cite{sachan2021identifying}) to each \textit{2H} data segment and identify the illicit DNs that have a $>$99.0\% probability of being malicious 
(computed as a ratio of the number of times a DN behaves maliciously and the total number of times the DN occurs). Then we identify the number of cryptojacked DNs present in our suspicious list identified in the first step. For the supervised ML models, we apply the reported supervised ML model (DecisionTree Classifier in~\cite{sachan2021identifying}) on \textit{ALL} data granularity to identify cryptojacked DNs. We also identify the best performing ML model along with the hyperparameters by configuring AutoML tools such as TPOT~\cite{olson2016tpot} with 11 different supervised ML algorithms with multiple combinations of their hyperparameters. We use unsupervised learning on \textit{2H} temporal granularity datasets as behavioral changes are better captured here than in the \textit{All} dataset. Applying unsupervised learning to \textit{All} dataset will only provide one class to the DNs, while in the other case, for each dataset in \textit{2H} granularity, we will get a class for each DN. As these classes would be associated with time, behavioral changes are captured over time.

To analyze the Indian Government websites, we crawl the list of URLs present in~\cite{goidirectory.gov.in}. We crawl the source code of the page behind the listed URL. Our crawler performs a signature search in the source code of the URL and all \textit{script} codes associated with the URL. If a signature is found in the source code during crawling, we mark/label it as cryptojacked. Then, we perform resources monitoring (using the ``\textit{iostat}'' Linux command) and capture the resource utilization for different URLs. Further, we use the PyShark wrapper to capture different measures from the network traces for each URL. Table~\ref{table:4} summarizes all list of resources we monitor. We do resource monitoring two times at an interval of 150 seconds and take the average of each measure we capture. 
We then analyze the collected data based on clustering and graph connectivity. We perform clustering to identify the DNs with distinct resource utilization and graph connectivity to analyze the association between the DNS records.

\begin{table}
\centering
\caption{Resources measures}
\label{table:4}
\begin{footnotesize}
\begin{tabular}{|c|l|}
    \hline
    Measures & Description \\
    \hline \hline
    cpu\_user & \% CPU used by user level applications\\
    cpu\_nice & \% CPU used by user level nice priority\\
    cpu\_system & \% CPU used by system level process\\
    cpu\_iowait & \% CPU idle time during which system had an\\ 
    &  outstanding disk I/O request\\
    cpu\_steal & \% time spent in involuntary wait by the virtual CPU\\
    cpu\_idle & \% time that CPU was idle and the system did
    not have \\ & an outstanding disk I/O request\\
    \hline
    sda\_tps & \# transfers per second that were issued to sda\\
    sda\_kB\_read/s & amount of blocks read/sec from sda \\ 
    sda\_kB\_wrtn/s & amount of blocks written/sec to sda \\
    sda\_kB\_read & \# blocks read \\
    sda\_kB\_wrtn & \# blocks written \\
    \hline
    disk\_read & disk reads\\
    disk\_write & disk writes\\
    \hline
    net\_recv & network receive \\
    net\_send & network send\\
    \hline
    pkt\_total & total packets\\
    pkt\_send & packets send \\
    pkt\_rec & packets received \\
    pkt\_oth & other packets \\
    \hline
\end{tabular}
\end{footnotesize}
\vspace{-0.4cm}
\end{table}

\section{Validation and Result Analysis}\label{sec:results}
We analyze the similarity between the cryptojacked DNs and malicious DNs and validate the effectiveness of the DN-based state-of-the-art such as~\cite{sachan2021identifying} to detect the cryptojacking DNs. We use Python and supporting libraries such as Beautiful Soup, Selenium Webdriver, PyShark, tldextract, whois, and DNS Resolver to build our approach. Our methodology is tested on a Linux machine with 1 TB storage, 64 GB RAM, and an I7 Intel core 3.2GHz CPU.

\subsection{Dataset}
Our approach is validated on the Cisco Umbrella top 1 million dataset~\cite{Umbrella1m} for January 2020. Our month choice is due to the limited computing power available to us and to keep the data the same as~\cite{sachan2021identifying}. It contains $\approx$335 million DNS queries. Of these, $\approx$1.77 million DNS queries are distinct, and 42002 DNS queries have the malicious tag (from~\cite{sachan2021identifying}). For ground truth on cryptojacking DNs, we use CoinHive BlackList~\cite{CoinHive-BlackList}, CoinHive Domains~\cite{CoinHive-Domains}, CoinHive Pixalate~\cite{CoinHive-Pixalate}, Cryptocurrency Mining List~\cite{Cryptocurrency-Mining-List}, Cryptojacking Campaign List~\cite{CryptoJacking-Campaign-List}, KnownCryptoURL~\cite{KnownCryptoURL}, MinerBlock List~\cite{MinerBlock}, NoCoin BlackList~\cite{NoCoinBlackList}, Top Web Mining Sites~\cite{TopWebMiningSites}, and the other websites such as~\cite{bijmans2019inadvertently}. We understand that some of these lists might be outdated, but we use them for the sake of completeness. There are 29777 unique cryptojacked DNs/TLDs (top-level domains) present in these lists. Only 1188 cryptojacked DNs are present in our dataset with corresponding 21743 DNS queries. Out of these 21743 DNS queries, 9681 DNS queries were unmarked previously and considered benign in~\cite{sachan2021identifying}. 

We use~\cite{goidirectory.gov.in} to get a list of 8669 Indian Government URLs as of $5^{th}$ August 2021 for our cryptojacking analysis. Out of these DNs, only 155 DNs are available in our dataset. Due to this limitation, we could not analyze the Indian Government websites using the considered Umbrella dataset. Thus, we perform analysis based on signature crawling, resource utilization, and associated DNS and whois records. We use a list of 66 cryptojacking signatures from studies such as~\cite{konoth2018minesweeper,ning2019capjack,bijmans2019inadvertently,saad2019dine,varlioglu2020cryptojacking} for signature crawling. 

\begin{figure*}
\hspace{-0.6cm}
    \centering
    \subfloat[Query Frequency]{
            \includegraphics[width=0.33\textwidth]{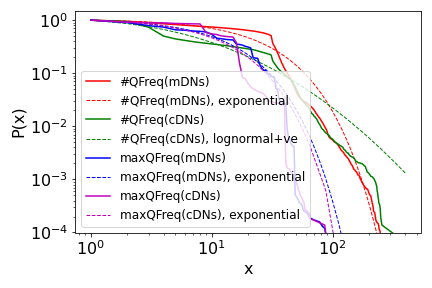}
        \label{fig:queryIE}
    }
    \subfloat[Query Frequency Bursts]{
            \includegraphics[width=0.33\textwidth]{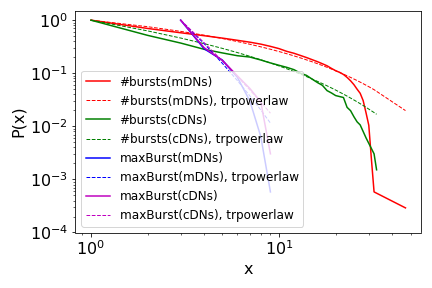}
        \label{fig:queryB}
    }
    \subfloat[Degree]{
            \includegraphics[width=0.33\textwidth]{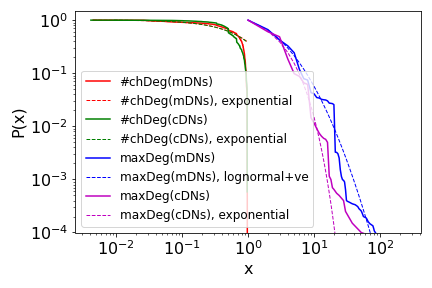}
        \label{fig:deg}
    }\\ 
    \hspace{-0.6cm}
    \subfloat[Diameter]{
            \includegraphics[width=0.33\textwidth]{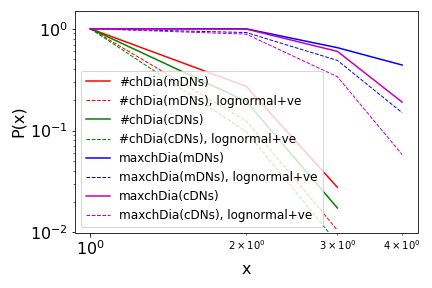}
        \label{fig:dia}
    }
    \subfloat[CPU utilization using \textit{iostat -c}] {
            \includegraphics[width=0.33\textwidth]{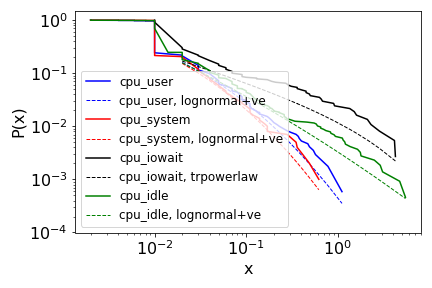}
        \label{fig:cpu}
    } 
        \subfloat[Device utilization using \textit{iostat -d sda}] {
            \includegraphics[width=0.33\textwidth]{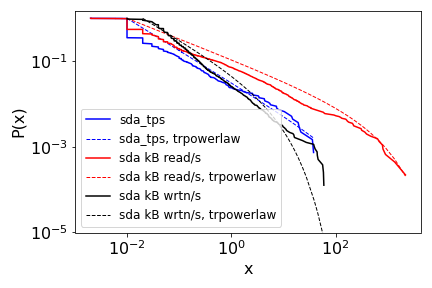}
        \label{fig:sda}
    }\\
    \hspace{-0.6cm}
    \subfloat[Disk utilization using \textit{dstat –disk}] {
            \includegraphics[width=0.33\textwidth]{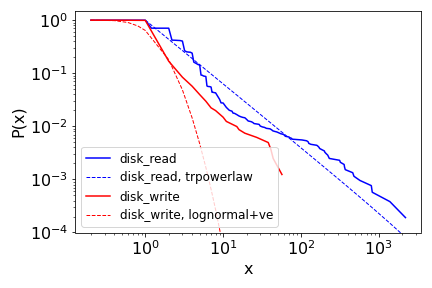}
        \label{fig:disk}
    }
    \subfloat[Network resources using \textit{pyshark}] {
            \includegraphics[width=0.33\textwidth]{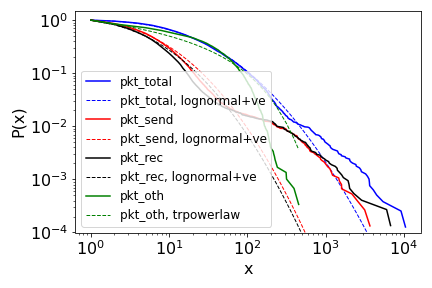}
        \label{fig:net}
    }
    \caption{Cumulative distribution of different   temporal properties and resource measures.}%
    \label{fig:analysis}%
    \vspace{-0.4cm}
\end{figure*}

\subsection{Similarity analysis between Cryptojacking DNs and Other Malicious DNs}

We compare the distribution of temporal properties such as query frequency, query frequency burst, degree, and diameter associated with cryptojacked DNs and other malicious DNs for the similarity analysis.
Here, we measure the behavioral similarity between the in-browser cryptojacked DNs (cDNs) and malicious DNs (mDNs) \textit{to decide whether a DN-based approach such as~\cite{sachan2021identifying}  can detect cryptojacking DNs.}
For this, we use \textit{ALL} granularity data of the Cisco Umbrella dataset.

First, we study the distribution of the number of query frequency \textit{(\#QFreq)} and the maximum query frequency \textit{(maxQFreq)} to analyze the similarity in \textit{query frequency}.   
Figure \ref{fig:queryIE} shows that the exponential distribution fits \#QFreq for the mDN class with $x_{min}$=1.0 and $\lambda$=0.0400, while a positive log-normal distribution fits the cDN class with $x_{min}$=1.0, $\mu$=1.7092, and $\sigma$=1.4067. As the two distributions are different, \textit{\#QFreq} may not be a good feature when detecting cryptojacking DNs using~\cite{sachan2021identifying}. Similarly, for the \textit{maxQFreq}, Figure \ref{fig:queryIE} shows that the exponential distribution fits for both classes with $x_{min}$={1.0} and $\lambda$=\{0.0793, 0.0852\}, respectively. The KL-Divergence (KLD) between the mDN and cDN class distributions is 0.0026. Thus, it reveals that \textit{maxQFreq} is a good feature to detect cryptojacking  using~\cite{sachan2021identifying}.

Next, we analyze the \textit{query frequency burst} (cf. Figure \ref{fig:queryB}). A query frequency burst is the frequency of a DNS query which is more than a predefined value (i.e., 80\% of the maximum number of DNS queries of a DN during a time frame). We compare the distributions of the number of query bursts \textit{(\#bursts)} and the maximum size of query burst \textit{(maxBurst)} for each mDN and cDN class. We observe that truncated-powerlaw best fits both classes and for both the features. For the \textit{\#bursts}, for the both classes, $x_{min}$=1.0, a cut-off parameters, $\beta$=$\frac{1}{\lambda}$ where $\lambda$=\{0.0415, 0.0426\}, and $\alpha$=\{1.0, 1.3878\}, respectively. Similarly, for the \textit{maxBurst}, for the both classes, $x_{min}$=3.0, $\alpha$=\{3.5325, 3.8962\}, and $\lambda$=\{0.2192, 0.1067\}, respectively. The KLD in the case of \textit{\#bursts} is 0.5244, while in the case of \textit{maxBurst} is 0.0746. This analysis also indicates that the two classes have the same statistical property with small divergence and will not impact the performance of~\cite{sachan2021identifying} when detecting cryptojacking DNs.

We next analyze the \textit{degree} (cf. Figure \ref{fig:deg}). Here, we compare the distributions of the number of times degree changes \textit{(\#chDeg)} and the maximum size of a degree \textit{(maxDeg)} over time for each class of DNs. Figure \ref{fig:deg} shows no valid fits for both classes out of exponential, positive log-normal, truncated-power-law, and power-law distributions for the \textit{\#chDeg}. Similarly, for the \textit{maxDeg}, Figure \ref{fig:deg} shows a positive log-normal distribution for mDN class with $x_{min}$=1.0, $\mu$=0.6831, and $\sigma$=0.9453, and exponential distribution for cDN class with $x_{min}$=1.0, and $\lambda$=0.4683. As the statistical properties are different here, \textit{maxDeg} can hamper the performance of~\cite{sachan2021identifying} when detecting cryptojacking DNs.

Next, we analyze the \textit{diameter} for both classes (cf. Figure \ref{fig:dia}). We study the distributions of the number of times diameter changes \textit{(\#chDia)} and maximum diameter change \textit{(maxchDia)}. We observe the positive log-normal distribution fits for both classes of DNs with $x_{min}$=1.0, $\mu$=\{0.1741, 0.0532\}, and $\sigma$=\{0.3773, 0.3986\} for the \textit{\#chDia}, respectively. The KLD between the mDN and cDN classes is 0.05. We also observe that positive log-normal distribution fits the best both class with $x_{min}$=1.0, $\mu$=\{1.0906, 0.9939\}, and $\sigma$=\{0.2856, 0.2499\} for the \textit{maxchDia}, respectively. The KLD between the distributions of both classes is 0.1036. 
This small divergence between the classes means that \textit{diameter} may not hamper the performance of~\cite{sachan2021identifying} when detecting cryptojacking DNs. 

From the above similarity analysis, we observe divergence in some features of the cDNs and mDNs.
Thus, the state-of-the-art feature vector (used for detecting the malicious DNs, i.e.,~\cite{sachan2021identifying}) can detect cryptojacking DNs, but some improvements are needed, and new features should be included.

\subsection{Effectiveness of a DN-based method~\cite{sachan2021identifying}}

\textit{To understand if there is an impact on the performance of state-of-the-art methods such as~\cite{sachan2021identifying} in identifying cryptojacked DNs/web pages}, we perform validations using reported unsupervised and supervised algorithms.

\subsubsection{Validation of an unsupervised model of~\cite{sachan2021identifying}} We apply K-Means (an unsupervised learning method) to each \textit{2H} data segment with different values of K$\in$[7,24]. This range of K is the same as identified in~\cite{sachan2021identifying}. The obtained results contain a series of labels for each granularity representing the number of times a particular DN showed malicious behavior. 
Among the 9681 cryptojacked DNs (those previously unmarked in the dataset), 9339 DNs show malicious behavior at least once. While only 228 DNs have the probability of being malicious $>$99.0\%. Now because we know the ground truth of these 228 DNs is cryptojacked, we can affirmatively say that the approach in~\cite{sachan2021identifying} is effective and is able to detect cryptojacked DNS.  As in this work, we do not propose any new feature, we do not quantify the effectiveness.

\subsubsection{Validation of the supervised model of~\cite{sachan2021identifying}} 
In our dataset, we have 9681 cryptojacked DNs with unmarked tags and 12062 cryptojacked DNs with malicious tags in the \textit{ALL} data granularity. It means that the reported supervised ML model (DecisionTrees Classifier) in~\cite{sachan2021identifying} is already trained with cryptojacked DNs. 
To validate the reported model for detecting the cryptojacked DNs, we perform an 80\%-20\% split of the dataset as well as the unmarked and cryptojacked DNs. The 80\% data is used for training while remaining for testing. This resized dataset has 9681 cryptojacked DNs and 186205 unmarked DNs (a total of 195886 DNs). 
We apply the DecisionTrees Classifier with the same hyperparameters, i.e., criterion=gini, max\_depth=10, min\_samples\_leaf=13, min\_samples\_split=12, splitter=best. Here, other hyperparameters have default values used by the Python scikit-learn library. It achieves 79.69\% \textit{Balance-Accuracy}. Here for cDN class  \textit{Precision} is 97.0\%, \textit{Recall} is  59.5\%, and  \textit{F1-score} is 74.0\%. Here, a low Recall on the cDN class signifies the need for improvement in the model~\cite{sachan2021identifying} for detecting cryptojacking DNs.  
These validations tests reveal that reported models in~\cite{sachan2021identifying} are able to detect the DNs which are involved in cryptojacking but with a low  Recall. Thus, next, we validate if there \textit{is any other supervised ML model that gives improved results?} 

\subsection{Identification of Improved ML Model}
To identify the supervised ML model that provides better results when identifying the cryptojacking DNs, we perform two tests using different data configurations (based on the distribution of cryptojacked DNs in the dataset). Here, we not only identify which supervised ML algorithm performs the best in our case but also report its hyperparameters. For this analysis, we use \textit{ALL} data granularity. To perform such validation, we use the AutoML tool called TPOT. We configure TPOT to use 11 different supervised algorithms with custom hyperparameters. These supervised algorithms are \textit{GaussianNB}, \textit{BernoulliNB}, \textit{DecisionTree}, \textit{RandomForest}, \textit{ExtraTrees}, \textit{K-NearestNeighbors}, \textit{GradientBoosting}, \textit{NeuralNetwork}, \textit{SupportVectorMachines}, \textit{LogisticRegression}, and \textit{EnsembleBagging}. TOPT reports the overall best-identified algorithm in terms of \textit{Balanced-Accuracy}. We also report \textit{Precision}, \textit{Recall}, and \textit{F1-score} for the cryptojacking class. 

We use two data configurations include: \textit{(i)} no cryptojacked DNs is present in the training dataset, and all cryptojacked DNs (i.e., 21743 DNs) are included in the testing dataset, and \textit{(ii)} cryptojacked DNs are distributed in an 80-20 ratio between training and testing data. 
TPOT reports \textit{DecisionTree} with different hyperparameters as the best classifier in both the test configurations. The results for both data configurations are listed in Table \ref{table:2}, along with the respective hyperparameters. The hyperparameters that have default values used by Python scikit-learn are  not reported here. Here, the results reveal a low \textit{Recall} on the cDN class. This is not better than the already reported DecisionTreee classifier in~\cite{sachan2021identifying}. Thus, it certainly signifies the need for improvements in the feature set of~\cite{sachan2021identifying} to better identify the cryptojacked DNs.

\begin{table}
\centering
\caption{Reported results by TPOT}
\label{table:2}
\begin{threeparttable}
\begin{tabular}{|c|c|c|c|c|c|c|}
    \hline
     \multicolumn{2}{|c|}{Cryptojacking} & \multirow{2}{*}{Classifier} & \multicolumn{4}{|c|}{\multirow{2}{*}{Results in (\%)}} \\
    \multicolumn{2}{|c|}{DNs in Dataset} & & \multicolumn{4}{c|}{} \\
    \hline 
     Train & Test & & BAcc & Pre & Rec & F1 \\
    \hline 
    - & 100\% & DT$\dagger$ & 67.56 & 86.0 & 35.64 & 50.0 \\
    \hline
    80\% & 20\% & DT$\ddagger$ & 72.02 & 85.0 & 44.45 & 58.0 \\
    \hline
    \multicolumn{2}{|c|}{Total} & 1771626 & \multicolumn{4}{c|}{} \\
    \hline
\end{tabular}
\begin{tablenotes}
        \item $\bullet$ \textbf{Train:} Training, \textbf{Test:} Testing, \textbf{DT:} DecisionTree, \textbf{BAcc:} Balance-Accuracy, \textbf{Pre:} Precision, \textbf{Rec:} Recall, \textbf{F1:} F1-score
        \item $\dagger$ criterion=gini, max\_depth=10, min\_samples\_leaf=13, min\_samples\_split=13, splitter=best,
        \item $\ddagger$ criterion=entropy, max\_depth=7, min\_samples\_leaf=18, min\_samples\_split=20, splitter=best. 
\end{tablenotes}
\end{threeparttable}
\vspace{-0.4cm}
\end{table}

\subsection{Analysis of Indian Government websites}
Indian Government websites are one of the preferred targets of cryptojackers because these websites have high traffic and end-user trust~\cite{reportET2018}. In the past, two Government websites of Andhra Pradesh, a state in India, have witnessed in-browser cryptojacking/cryptomining activities~\cite{reportET2018}.
\textit{To identify the Indian Government websites that are compromised for cryptocurrency mining}, we perform our analysis in three parts, \textit{(i)} based on the signature crawling, \textit{(ii)} based on resource utilization, i.e.,  CPU,  Device,  Disk, and Network,  and \textit{(iii)} based on the association between the DNS records of websites (i.e., DN, associated IP addresses, Name-Server, and Country).

With signature crawling, we identify the cryptojacking DNs based on the past reported signatures and mark them as suspicious for further analysis. 
The crawler opens each webpage associated with a DN using \textit{selenium webdriver}, reads it, and searches the existence of  66 cryptojacking signatures in its \textit{HTML code} and all associated \textit{script codes}. Here, against each webpage, we record the matching \textit{signatures} present on the webpage. We identify 47 webpages have \textit{monero} keyword, and only 1 URL has a \textit{coin} keyword. However, none-of-them are associated with cryptomining. All the \textit{monero} keywords are associated with the font family, and the \textit{coin} is linked with a widget. This analysis shows that none-of-the Indian webpages currently contain the cryptojacking signature in their code during the mentioned period. 

Next, we analyze resource utilization for each webpage using \textit{iostat-c} to measure the \textit{CPU utilization} (cf. Figure \ref{fig:cpu}), \textit{iostat-d sda} to measure the \textit{device utilization} (cf. Figure \ref{fig:sda}), and \textit{dstat-disk-net} to measure the \textit{disk utilization} statistics (cf. Figure \ref{fig:disk}). We also use the \textit{PyShark} wrapper to analyze live network packets (cf. Figure \ref{fig:net}). We set the \textit{PyShark} timeout to 30 sec, the \textit{selenium webdriver} timeout to 90 sec, and the time gap between two resource measuring points to 150 sec for resource and network analysis. We perform this analysis from November to December 2021 and record the 19 resource measures (cf. Table \ref{table:4}). From Figure \ref{fig:cpu}, we infer that the truncated-powerlaw best fits \textit{cpu\_iowait} with $\alpha$=1.54 and $\lambda$=0.118 and the positive log-normal fits for \textit{cpu\_user}, \textit{cpu\_system}, and \textit{cpu\_idle} with $\mu$=\{-5.08, -5.36, -14.24\}, $\sigma$=\{1.57, 1.53, 3.65\}, respectively. Here, $x_{min}$ is 0.015 for the all four CPU measures. Similarly, from Figure \ref{fig:sda}, we infer that the positive log-normal distribution fits \textit{sda\_tps}, \textit{sda\_kB\_read/s}, and \textit{sda\_kB\_wrtn/s} with $x_{min}$=\{0.01, 0.01, 0.02\}, $\alpha$=\{1.75, 1.47, 1.71\} and $\lambda$=\{0.001, 0.0005, 0.06\}, respectively. 
Next, Figure \ref{fig:disk} shows truncated-powerlaw best fits \textit{disk\_read} with $x_{min}$=1.0, $\alpha$=2.20 and $\lambda$=2.72 and a positive log-normal distribution best fits \textit{disk\_write} with $x_{min}$=0.1, $\mu$=0.19, $\sigma$=0.54. 
Similarly, Figure \ref{fig:net} shows truncated-powerlaw best fits \textit{pkt\_oth} with $x_{min}$=1.0, $\alpha$=1.0 and $\lambda$=0.006 and the positive log-normal fits best for \textit{pkt\_total}, \textit{pkt\_send}, and \textit{pkt\_rec} with $x_{min}$=1.0, $\mu$=\{2.82, 1.58, 1.37\}, $\sigma$=\{1.42, 1.26, 1.28\}, respectively.
  

Further, we apply the \textit{K-Means} algorithm to the entire recorded dataset to cluster the DNs with K$\in$[2, 15]. Our choice (range on K) is based on the data size. We choose the best K based on the silhouette score. We check the silhouette score for different values of K and find K=2 provides the best silhouette score of 0.975 (different silhouette scores obtained for different values of K are listed in Table~\ref{table:3}). After exploring the clusters obtained for K=2, we find that one cluster has 8624 DNs while the second cluster has only 10 DNs, indicating that these 10 DNs have different properties than the others and should be monitored. Note that we do not use supervised ML algorithms such as \textit{DecisionTree} for the analysis due to the unavailability of the ground truth of Indian Government websites.

\begin{table}
    \centering
    \caption{Silhouette Scores (S)} 
    \label{table:3}
    \begin{tabular}{|c|c|c|c|c|c|c|c|}
        \hline
         \textbf{K} & 2 & 3 & 4 & 5 & 6 & 7 & 8 \\
         \hline
         \textbf{S} & \textbf{0.97} & 0.95 & 0.94 & 0.72 & 0.70 & 0.65 & 0.67\\
         \hline \hline
         \textbf{K} & 9 & 10 & 11 & 12 & 13 & 14 & 15\\
         \hline
          \textbf{S} & 0.66 & 0.66& 0.66 & 0.39 & 0.34 & 0.32 & 0.23\\
         
         \hline
    \end{tabular}
    \vspace{-0.4cm}
\end{table}

After the signature crawling and resource utilization, we extract features for each DN using the \textit{whois} and \textit{tldextract}. These features are: \textit{subdomain, registered\_domain, creation\_date, updated\_date, age, last\_updated\_age, Country, A Record (IPv4 Address record), AAAA Record (IPv6 Address record), NS (Name Server), MX (Mail Exchanger), TXT (Text), CNAME (Canonical Name), DNAME (Delegation Name), SOA (Start of Authority)}. Next, we build a graph using IP and NS addresses. We identify 7 connected components in the graph, and 8658 out of 8669 webpages lie in the largest component, where all the DNs are hosted on National Informatics Center servers. We also find that 1839 DNs do not have the \textit{Country} entry, and the remaining DNs have 21 unique countries. Out of these, DNs of 6728 webpages are hosted in India, DNs of 48 webpages are hosted in the USA, and DNs of 10 webpages are hosted in Estonia. We find one DN each is hosted in countries such as Iceland, Canada, United Kingdom, Singapore, Netherlands, Belize, China, Hong Kong, Hong Kong, Indonesia, Ukraine, Romania, Japan, Panama, Brazil, Belarus, France, and Switzerland.
\vspace{-0.2cm}
\section{Conclusion}\label{sec:conc}
Detection of in-browser cryptojacking is essential to safeguard users' systems from illegal mining activities. Past approaches have used various techniques to detect in-browser cryptojacking, such as signature crawling and resource analysis. Besides these techniques, meta-information attached with a domain name also provides valuable inputs to detect cryptojacking. In this work, we validate a metadata-based technique~\cite{sachan2021identifying} to detect the in-browser cryptojacking DNs. This technique uses metadata information of DNs and associated temporal and non-temporal properties for malicious DNs detection. We also perform a comparative study of various techniques that detect in-browser cryptojacking DNs.

Our analysis shows behavior similarity exists between the cryptojacking DNs and other suspicious DNs. At the same time, there is a need for improvement in the feature set of~\cite{sachan2021identifying} to improve the results of the approach. Our signature-based analysis also identifies that none-of-the Indian Government websites listed in~\cite{goidirectory.gov.in} were involved in in-browser cryptojacking from October-December 2021. Our resource utilization analysis finds different resource utilization by 10 DNs. Such DNs require continuous and detailed behavior analysis before marking them as suspects. Finally, we conclude that we need to enhance the feature set of the metadata-based approach with resources and network analysis-based features.

In the future, we would like to improve the metadata-based approach and test it in a large dataset to detect in-browser cryptojacking. We would also like to develop temporal data of Indian Government websites, which will be helpful for the metadata-based approach in the future.
\vspace{-0.1cm}
\section*{Acknowledgement}\label{sec:ack}
This work is partially funded by the National Blockchain Project at IIT Kanpur, sponsored by the National Cyber Security Coordinator's office of the Government of India, and partially by the C3i Hub funding from the Department of Science and Technology of the Government of India. We also thank Hugo~L.~J.~Bijmans, for making the cryptojacking dataset available to us.

\vspace{-0.2cm}
\bibliographystyle{ieeetr}
\bibliography{biblio.bib}

\end{document}